\documentclass[preprint,aps,12pt,showpacs,nofootinbib,tightenlines]{revtex4}
\usepackage{amsmath}
\usepackage{amssymb}
\usepackage{CJK}
\usepackage{epsfig}
\usepackage{graphicx}
\usepackage{subfigure}
\usepackage{slashed}
\begin{document}
\def\pslash{\rlap{\hspace{0.02cm}/}{p}}
\def\eslash{\rlap{\hspace{0.02cm}/}{e}}
\title {Top Quark FCNC Decay to 125GeV Higgs boson \\in the Littlest Higgs Model with T-parity}
\author{Bingfang Yang$^{1,2,3}$}
\author{Ning Liu$^{1,4}$}\email{wlln@mail.ustc.edu.cn}
\author{Jinzhong Han$^{5}$}
\affiliation{ $^1$College of Physics $\&$ Electronic Engineering,
Henan Normal University, Xinxiang 453007, China\\
$^2$School of Materials Science and Engineering, Henan Polytechnic
University, Jiaozuo 454000, China\\
$^3$ Basic Teaching Department, Jiaozuo University, Jiaozuo 454000, China\\
$^4$ ARC Centre of Excellence for Particle Physics at the Terascale,
School of Physics, The University of Sydney, NSW 2006, Australia\\
$^5$Department of Physics and Electronic Engineering, Zhoukou Normal
University, Henan, 466001, China
   \vspace*{1.5cm}  }

\begin{abstract}

Motivated by the current observation of a 125 GeV Higgs boson, we
calculate $t\rightarrow cH$ and $t\rightarrow cg(\gamma)H$ in the
unitary gauge in the littlest Higgs model with T-parity(LHT). Due to
the large contribution from the new mirror fermions, we find that
the branching ratios of $t\rightarrow cH$ and $t\rightarrow cgH$ can
be greatly enhanced in the LHT model and maximally reach $\mathcal
O(10^{-5})$ in the allowed parameter space. When the mirror fermion
mass $M_3 > 2(1.5)$ TeV and the cut-off scale $f=500$ GeV, the
process of $pp \to t\bar{t} \to 3b +c +\ell+ \slashed E_{T}^{miss}$
can reach $3\sigma$($5\sigma$) sensitivity at 8(14) TeV LHC with
luminosity $\cal{L}$ $=20(300) fb^{-1}$.

\end{abstract}
 \maketitle
\section{ Introduction}
\noindent

In the light of the discovery of a Standard Model(SM) like Higgs
boson and the null results of new physics at the LHC, the
electroweak hierarchy problem is highlighted much more than ever
before. As the heaviest known elementary particle, top quark has a
strong correlation with the hierarchy problem and can be identified
as a smoking gun of the TeV-scale physics.

In the SM, the top quark flavor-changing neutral-current(FCNC)
processes are highly suppressed by the
Glashow-Iliopoulos-Maiani(G.I.M.) mechanism\cite{GIM}. It indicates
that any observation of these processes will be a signal beyond the
SM\cite{FCNCdecay,aguila,fajfer,ferreira,other}. Since weakly
constrained FCNC couplings between the second and the third
generation up-type quarks are usually predicted in some new physics
models, the two-body FCNC decays $t\rightarrow cX(X=g,\gamma,Z,H)$
can be greatly enhanced, such as in the minimal supersymmetric
standard model (MSSM) with branching radio Br($t\rightarrow cH)\sim
10^{-5}$\cite{FCNCMSSM}, R-parity violating SUSY with branching
fraction Br($t\rightarrow cH)\sim 10^{-6}$\cite{RPVtch}, the
two-Higgs doublet model (2HDM) with branching radio Br($t\rightarrow
cH)\sim 10^{-3}$\cite{FCNC2HDM} and so on. The NLO QCD corrections to $t \to qH(q=u,c)$ in a model-independent method
has been studied in the Ref\cite{fabio}.
Besides, three-body FCNC decays of the top quark were also found to
be a sensitive probe of the new physics, such as $t\rightarrow
cX_{1}X_{2}(X=g,\gamma,Z,H)$\cite{2FCNC2HDM,2FCNCMSSM,2FCNCLHT}. Very recently,
ATLAS collaboration has measured the top quark decays $t \to cH$ with $H \to \gamma\gamma$ and set the upper limit on the $tcH$ coupling as 0.17\cite{atlas-tch}.

As an extension of the SM, the Littlest Higgs Model(LHT)\cite{LHT}
model can successfully solve the electroweak hierarchy problem by
constructing the Higgs as a Pseudogoldstone boson. Meanwhile, the
discrete symmetry T-parity in this model also forbids the tree-level
contributions from the heavy gauge bosons, thus LHT can safely avoid
the constraint from the electroweak precision obeservables(EPO) that
occurs in the littlest Higgs (LH) model \cite{LH}. For top quark
sector in the LHT, the top quark can interact with new T-odd gauge
bosons and T-odd fermions, which may produce large contributions to
the top quark FCNC processes\cite{topFCNC-LHT}. The similar effects
have been studied in the rare decays of $K$/$B$-meson\cite{KB},
Higgs boson\cite{HZ} and $Z$ boson\cite{HZ}.

It should be mentioned that the searches for the LHT particles at
the LHC can provide the direct evidence of the LHT model or give a
strong constraints on the LHT parameter space. However, the results
usually depend on the assumption of the specific mass spectrum and
the branching ratios. For example, the T-odd top partner($T^{-}$)
pair production has been explored through $pp \to T^{-}T^{-} \to A_H
t A_H \bar{t}$ at 7 TeV LHC\cite{ttprl}. In the analysis, a large
mass splitting between the $A_H$ and $T^{-}$ is required to produce
the hard missing energy to suppress the top pair background. But in
a general LHT model, the mass of $T^{-}$ can be close to the mass of
$(A_H + t)$ or $A_H$ so that the adopted strategy is not applicable.
The similar things can also happen in the searches for other LHT new
particles. So in these cases, the searches for the indirect LHT
effects via loop corrections will be of great importance because of
its weak dependence on the kinematics information. In particular,
the processes with low SM backgrounds, such as top quark FCNC
decays, will be helpful for testing the LHT model.

In this work, we calculate the top quark FCNC decays with Higgs
interactions in unitary gauge in the LHT, that is, $t\rightarrow cH$
and $t\rightarrow cg(\gamma)H$. As a top quark factory, 14 TeV LHC
has a power in detecting the branching ratios of $t\to cH$ up to $Br
\sim \mathcal O(10^{-6})$ for $\mathcal{L}=30fb^{-1}$ and $Br \sim
\mathcal O(10^{-7})$ for $\mathcal{L}=300fb^{-1}$\cite{FCNCLHC}. So
the study of these top FCNC processes can be used to test the LHT at
the LHC. The paper is organized as follows. In Sec.II we
recapitulate the LHT model related to our work. In Secs.III and
Secs.IV we calculate the one-loop contributions of the LHT model to
the $t\rightarrow cH$ and $t\rightarrow cg(\gamma)H$ in unitary
gauge and present the numerical results. Finally, we give a short
summary in Sec.V.

\section{ A brief review of the LHT model}
\noindent The LHT model is a non-linear $\sigma$ model based on the
coset space $SU(5)/SO(5)$, with the $SU(5)$ global symmetry broken
by the vacuum expectation value (VEV) of a $5 \times 5$ symmetric
tensor,
\begin{eqnarray}
\Sigma_0=
\begin{pmatrix}
{\bf 0}_{2\times2} & 0 & {\bf 1}_{2\times2} \\
                         0 & 1 &0 \\
                         {\bf 1}_{2\times2} & 0 & {\bf 0}_{2\times 2}.
\end{pmatrix}
\end{eqnarray}

The VEV of $\Sigma_0$ breaks the extended gauge group $\left[ SU(2)
  \times U(1) \right]^2$ down to the SM electroweak
  $SU(2)_L \times U(1)_Y$, which leads to new heavy gauge bosons
$W_{H}^{\pm},Z_{H},A_{H}$ with the masses given to lowest order in
$v/f$ by
\begin {equation}
M_{W_{H}}=M_{Z_{H}}=gf(1-\frac{\upsilon^{2}}{8f^{2}}),~~M_{A_{H}}=\frac{g'f}{\sqrt{5}}
(1-\frac{5\upsilon^{2}}{8f^{2}})
\end {equation}
Here $g$ and $g'$ are the SM $SU(2)$ and $U(1)$ gauge couplings,
respectively.

When T-parity is implemented in the quark sector of the model, we
require the existence of mirror partners with T-odd quantum number
for each SM quark. We denote them by $u_{H}^{i},d_{H}^{i}$, where
$i$($i=1,2,3$) is the generation index. After electroweak symmetry
breaking (EWSB), a small mass splitting between $u_{H}^{i}$ and
$d_{H}^{i}$ is induced, and the masses are given by
\begin{equation}
m_{d_{H}^{i}}=\sqrt{2}\kappa_if, ~~m_{u_{H}^{i}}=
m_{d_{H}^{i}}(1-\frac{\upsilon^2}{8f^2})
\end{equation}
where $\kappa_i$ are the diagonalized Yukawa couplings of the mirror
quarks.

In order to stabilize the Higgs mass, an additional T-even heavy
quark $T^{+}$ is introduced to cancel the large one-loop quadratic
divergences caused by the top quark. But the implementation of
T-parity requires a T-odd mirror partner $T^{-}$ with $T^{+}$. Their
masses are given by
\begin{eqnarray}
m_{T^{+}}&=&\frac{f}{v}\frac{m_{t}}{\sqrt{x_{L}(1-x_{L})}}[1+\frac{v^{2}}{f^{2}}(\frac{1}{3}-x_{L}(1-x_{L}))]\\
m_{T^{-}}&=&\frac{f}{v}\frac{m_{t}}{\sqrt{x_{L}}}[1+\frac{v^{2}}{f^{2}}(\frac{1}{3}-\frac{1}{2}x_{L}(1-x_{L}))]
\end{eqnarray}
where $x_{L}$ is the mixing parameter between the SM top-quark and
its heavy partner $T^{+}$.

In the LHT model, the mirror quark Yukawa interaction is given by
\begin{eqnarray}
\mathcal{L}_{mirror}=-\kappa_{ij}f\left(\bar\Psi_2^i\xi +
  \bar\Psi_1^i\Sigma_0\Omega\xi^\dagger\Omega\right)\Psi_R^j+h.c.\
\end{eqnarray}

A new flavor structure can come from the mirror fermions when the
mass matrix $\sqrt{2}\kappa_{ij}f$ is diagonalized by two $U(3)$
matrices. One of the important ingredients of the mirror quark
sector is the existence of two CKM-like unitary mixing matrices:
$V_{Hu},V_{Hd}$. These mirror mixing matrices parameterize flavor
changing interactions between SM quarks and mirror quarks that are
mediated by the heavy gauge bosons $W_{H}^{\pm},Z_{H},A_{H}$.

Note that $V_{Hu}$ and $V_{Hd}$ are related through the SM CKM
matrix:
\begin{equation}
V_{Hu}^{\dag}V_{Hd}=V_{CKM}.
\end{equation}
We follow Ref.\cite{vhd} to parameterize $V_{Hd}$ with three angles
$\theta^d_{12},\theta^d_{23},\theta^d_{13}$ and three phases
$\delta^d_{12},\delta^d_{23},\delta^d_{13}$
\begin{eqnarray}
V_{Hd}=
\begin{pmatrix}
c^d_{12}c^d_{13}&s^d_{12}c^d_{13}e^{-i\delta^d_{12}}&s^d_{13}e^{-i\delta^d_{13}}\\
-s^d_{12}c^d_{23}e^{i\delta^d_{12}}-c^d_{12}s^d_{23}s^d_{13}e^{i(\delta^d_{13}-\delta^d_{23})}&
c^d_{12}c^d_{23}-s^d_{12}s^d_{23}s^d_{13}e^{i(\delta^d_{13}-\delta^d_{12}-\delta^d_{23})}&
s^d_{23}c^d_{13}e^{-i\delta^d_{23}}\\
s^d_{12}s^d_{23}e^{i(\delta^d_{12}+\delta^d_{23})}-c^d_{12}c^d_{23}s^d_{13}e^{i\delta^d_{13}}&
-c^d_{12}s^d_{23}e^{i\delta^d_{23}}-s^d_{12}c^d_{23}s^d_{13}e^{i(\delta^d_{13}-\delta^d_{12})}&
c^d_{23}c^d_{13}
\end{pmatrix}
\end{eqnarray}
In our calculation, for the matrices $V_{Hu},V_{Hd}$, to aid
comparisons with Ref.\cite{houhongsheng}, we also follow
Ref.\cite{case} to consider the same scenarios as follows

\begin{itemize}
\item {Case I,}
$V_{Hd} = {\mathbf 1}$
\item {Case II,}
 $s^d_{12}=\frac{1}{\sqrt{2}},
 ~s^d_{23}=5\times 10^{-5},
 ~s^d_{13}=4\times 10^{-2},
 ~\delta^d_{12}=\delta^d_{23}=\delta^d_{13}=0$
\item {Case III,}
 $s^d_{12}=0.99,
 ~ s^d_{23} = 2\times10^{-4},
 ~ s^d_{13} = 0.6,
 ~ \delta^d_{12}=\delta^d_{23}=\delta^d_{13}=0$
\end{itemize}
\section{Branching ratio for $t\rightarrow cH$ in the LHT model}

\noindent

In the LHT model, the relevant Feynman diagrams of the the process
$t\rightarrow cH$ in unitary gauge are shown in Fig.1. We can see
there is no additional mixing between $T^{+}$ and charm or up quark.
This is different from the case in Ref.\cite{exotic quarks}, where a
small loop induced coupling between new vector-like quark and charm
quark can occur and will be constrained by the low energy physics.
We will not consider the higher order couplings between the scalar
triplet $\Phi$ and top quark and neglect the high order $\mathcal
O(\upsilon^{2}/f^{2})$ terms in the masses of new particles. The
calculations of the loop diagrams are straightforward. We adopt the
definitions of scalar one-loop integral functions in
Ref.\cite{scalar} and compose each loop diagram into some scalar
loop functions \cite{loop function}, and list the explicit
expressions of these amplitudes in Appendix. We use the package of
LOOPTOOLS\cite{loop tools} to perform the numerical loop
calculations.  In the analytic calculations, we cancel the
divergence that is independent on the mirror quark mass by the
unitarity of the matrix $V_{Hu}$. We also note that the modified
interactions of the up-type mirror fermions with the $Z$ boson in
Ref.\cite{T. Goto} can cancel the similar divergence in the
processes with down-type quarks or leptons as the external
particles. However, we checked that such a modification cannot
cancel the divergence in $t\rightarrow cH$ and there is still
so-called left-over divergence\cite{left-over,houhongsheng} as
follow,
\begin{eqnarray}
D=\frac{m^{2}_{u^{i}_{H}}}{f^{2}}(V_{Hu})^{\ast}_{i2}(V_{Hu})_{i3}\frac{i}{16\pi^{2}}[-\frac{1}{80}+(\frac{x_{L}^{2}}{160}+\frac{3}{64})\frac{v^{2}}{f^{2}}]\Delta
\end{eqnarray}
where $\Delta=\frac{1}{\varepsilon}-\gamma_{E}+ln4\pi$. In
Ref.\cite{left-over}, this so-called left-over divergence was
understood as the sensitivity of the decay amplitudes to the
ultraviolet completion of the LHT model. Follow
Ref.\cite{left-over}, we remove the divergent term $1/\varepsilon$
in the invariant amplitudes and take the renormalization scale $\mu
= \Lambda$ with $\Lambda = 4\pi f$ being the cutoff scale of the LHT
model.

\begin{figure}[htbp]
\scalebox{0.4}{\epsfig{file=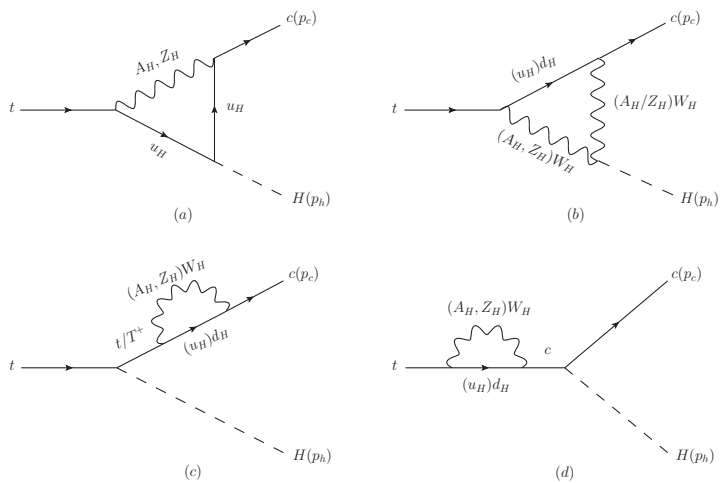}}\vspace{-0.5cm}\caption{Feynman
diagrams of the LHT one-loop correction to $t\rightarrow cH$ in the
unitary gauge.}
\end{figure}

In our numerical calculations, the SM parameters are taken as
follows\cite{parameters}
\begin{eqnarray}
\nonumber &&G_{F}=1.16637\times 10^{-5}GeV^{-2},
~\sin^{2}\theta_{W}=0.231,~\alpha_{e}=1/128,~\alpha_{s}=0.1076,\\
&&m_{c}=1.27GeV,~M_{Z}=91.1876GeV,~m_{t}=173.5GeV,~m_{h}=125GeV.
\end{eqnarray}

The relevant LHT parameters in our calculation are the scale $f$,
the mixing parameter $x_{L}$, the mirror quark masses and the
parameters in the matrices $V_{Hu},V_{Hd}$. Considering the
constraints in Refs.\cite{constraints}, the scale $f$ can be allowed
as low as 500 GeV. For the mirror quark masses, it has been shown
that the experimental bounds on four-fermi interactions require
$m_{Hi}\leq4.8f^{2}/\rm {TeV}$\cite{massbound}, we get
$m_{u_{H}^{i}}=m_{d_{H}^{i}}$ at $\mathcal O(\upsilon/f)$ and
further assume
\begin{equation}
m_{u_{H}^{1}}=m_{u_{H}^{2}}=m_{d_{H}^{1}}=m_{d_{H}^{2}}=M_{12},m_{u_{H}^{3}}=m_{d_{H}^{3}}=M_{3}
\end{equation}

\begin{figure}[htbp]
\begin{center}
\scalebox{0.7}{\epsfig{file=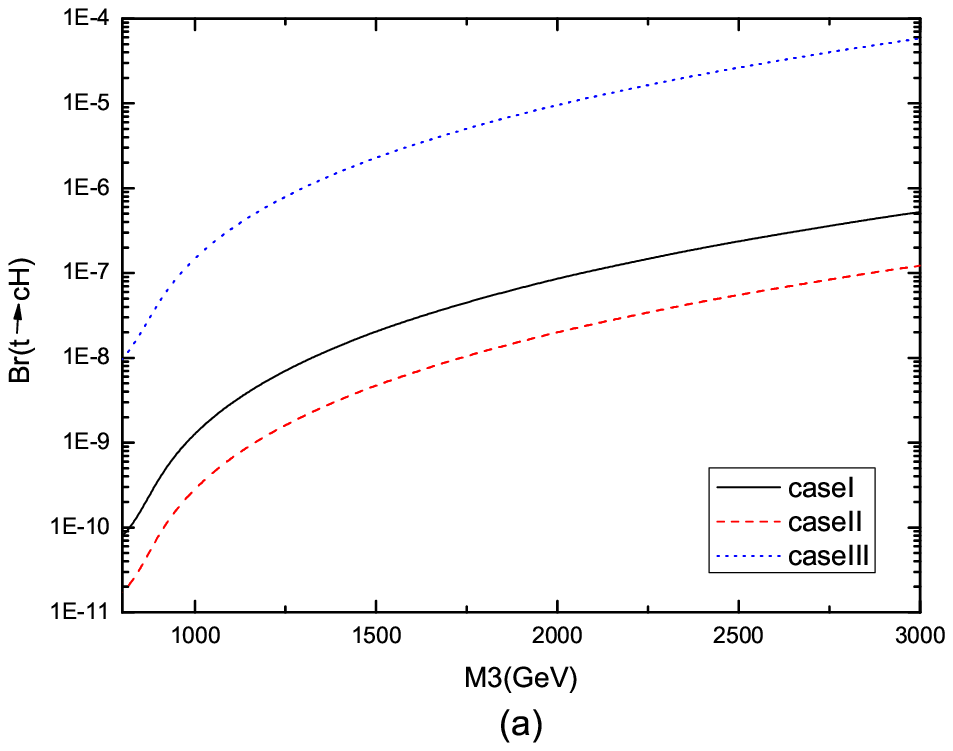}}\vspace{-1cm}\hspace{-1cm}
\scalebox{0.7}{\epsfig{file=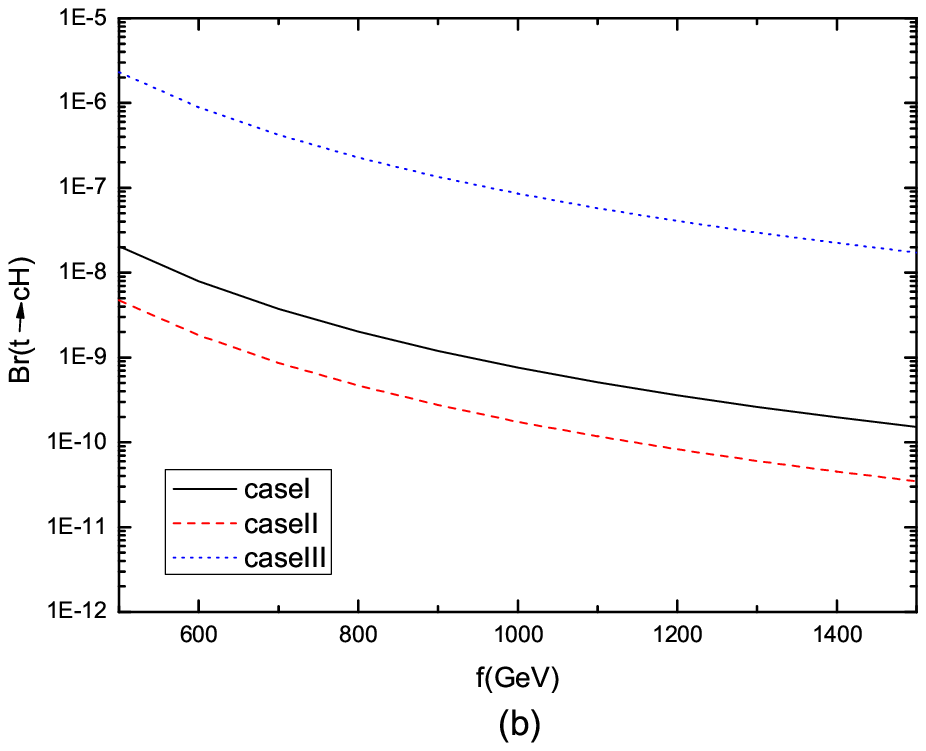}} \caption{Branching ratios of
$t\rightarrow cH$ as a function of $M_{3}$ (a) and $f$ (b) in three
cases, respectively.}
\end{center}
\end{figure}

In Fig.2(a), we show the dependance of the branching ratio of
$t\rightarrow cH$ on the third generation mirror quark mass $M_{3}$.
We set the scale $f=500$ GeV, the mixing parameter $x_{L}=0.1$ and
the first two generation mirror quark masses $M_{12}=750$ GeV. Due
to the large departures from the SM caused by mixing matrice in case
III, we can also see that the branching ratio of $t \to cH$ in the
case III is much larger than case I and II, which can maximally
reach $5.8\times 10^{-5}$ in case III.

From the Fig.2(a), we can see that the branching ratio of
$t\rightarrow cH$ increases with $M_{3}$ increasing, which means
that the decay rate is enhanced by the mass splitting between the
three generation mirror quarks. According to the analytic
expression, we can know the form factors of the loop-induced $tcH$
interaction, $F$, should take the following form
\begin{eqnarray}
F &\propto& \sum_{i=1}^3 \left ({V^\dag_{Hu}}\right)_{ci} f(m_{Hi})
\left({V_{Hu}}\right)_{it}
\end{eqnarray}
where $f(m_{Hi})$ is a universal function for three generation
mirror quarks, but its value depends on the mass of $i$th-generation
mirror quark, $m_{Hi}$. Obviously, for the degeneracy of the three
generation mirror quarks, $F$ vanished exactly due to the unitary of
$V_{Hu}$, while for the degeneracy of the first two generations as
discussed below, the factor behaviors like  $( V^\dag_{Hu} )_{c3} (
f(m_{H3}) - f(m_H) ) ({V_{Hu}})_{3t} $ with $m_H$ being the common
mass of the first two generations. The decay rate is enhanced by the
mass splitting between the three generation mirror quarks, since we
set $M_{1} = M_{2}=M_{12}$, there is only one mass splitting
$M_{3}-M_{12}$, which increases with $M_{3}$ while keeping $M_{12}$
fixed. This agrees with the explanation in Ref.\cite{topFCNC-LHT}.

In Fig.2(b), we show the dependance of the branching ratio of
$t\rightarrow cH$ on the scale $f$. We set the mixing parameter
$x_{L}=0.1$, the first two generation mirror quark masses
$M_{12}=1.5f$ and the third generation mirror quark mass $M_{3}=3f$.
We can see that the branching ratio decreases with the scale $f$
increasing, which means that the correction of the LHT model
decouples with the scale $f$ increasing. Since the enhancement from
mass splitting of mirror fermions can balance the suppression of
large scale $f$, we can find that the branching ratio of
$t\rightarrow cH$ decreases slowly, when the scale $f$ becomes
higher. From Fig.2, we can see that the LHT model can enhances the
branching ratios of $t\rightarrow cH$ as much as $9\sim10$ orders of
the one in the SM\cite{SMtch}. Similarly, in some other new physics
beyond the SM this branching ratio can also be enhanced by several
orders of magnitude. For comparison, we summarize the FCNC decays
$t\rightarrow cH$ in the LHT model and in other new physics
models\cite{top-FCNC,FC2HDM,MSSMtch,RPVtch,SUSYtch} in Table I.

\begin{figure}[htbp]
\begin{center}
\scalebox{0.7}{\epsfig{file=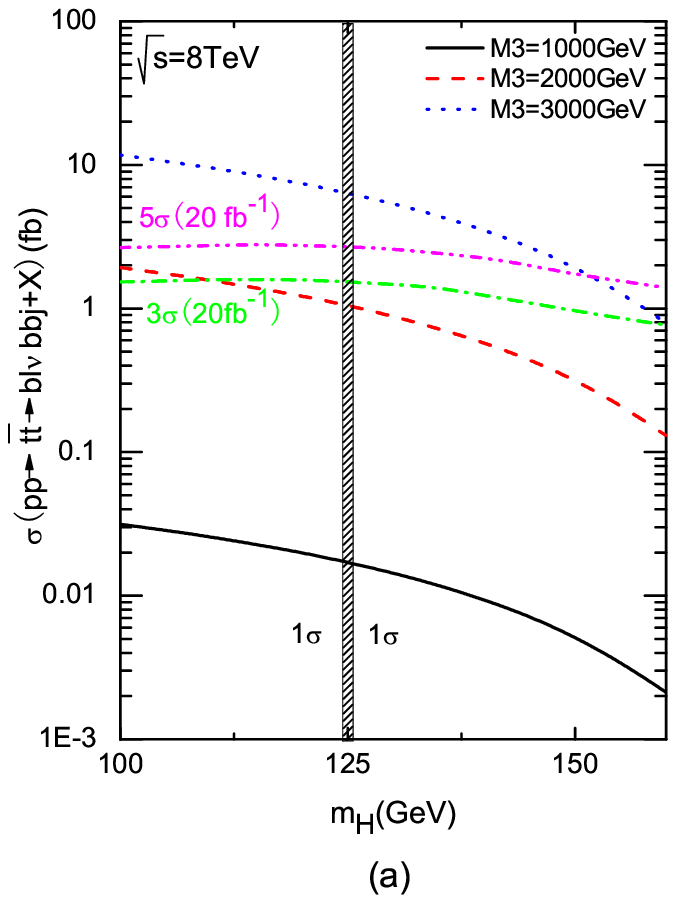}}\hspace{-1cm}
\scalebox{0.7}{\epsfig{file=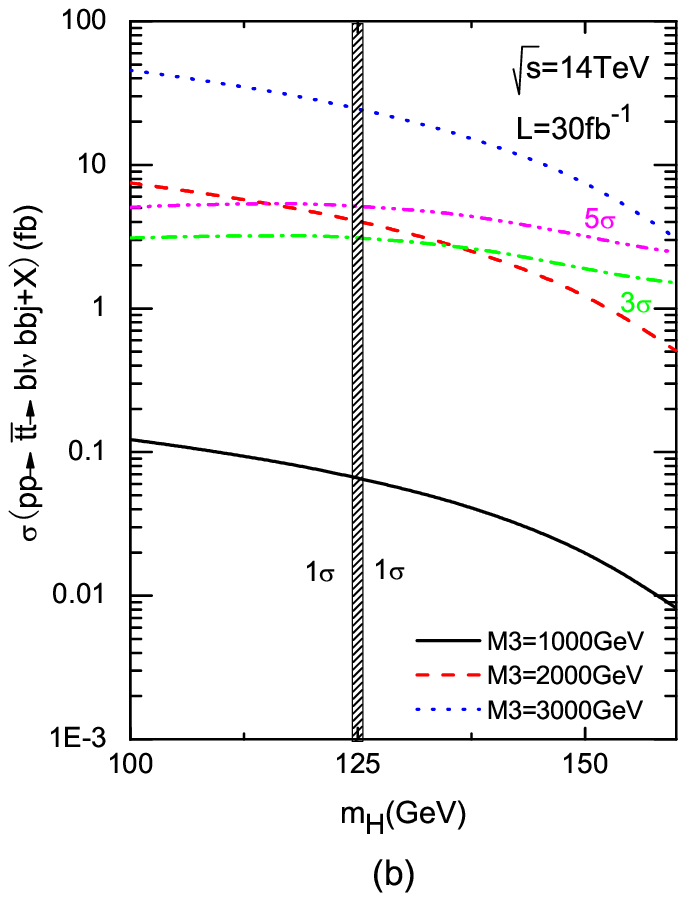}}\vspace{-1cm}\hspace{-1cm}
\scalebox{0.7}{\epsfig{file=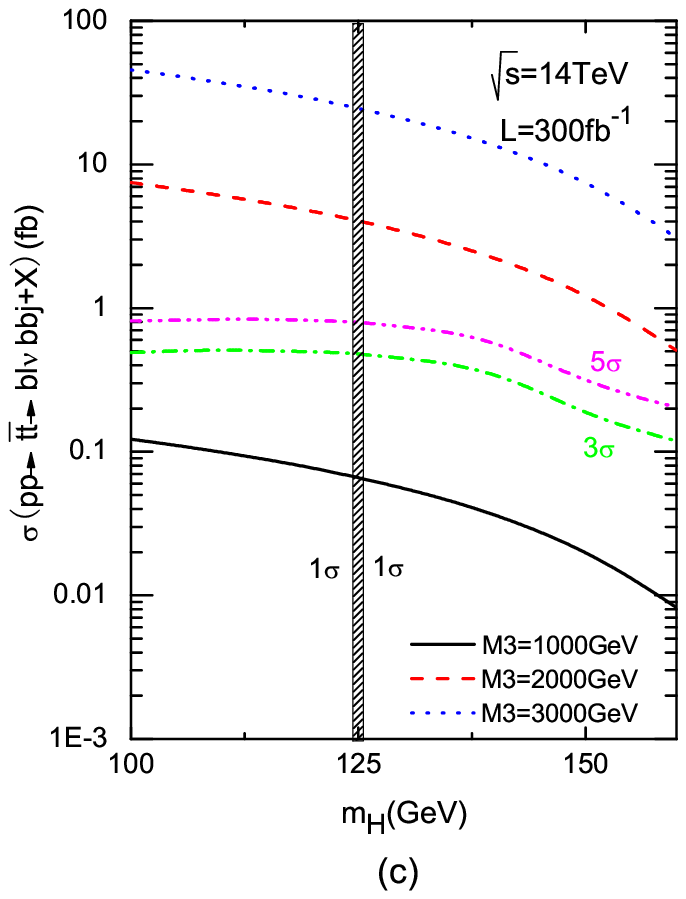}} \caption{The observability
of $t \to cH$ for case III in the LHT model through the production
of $pp\rightarrow t\bar{t}\rightarrow tcH\rightarrow bl\nu
cb\bar{b}+X$ at the LHC with $\sqrt{s}=8$ TeV and $\sqrt{s}=14$ TeV.
The shadow region is $1\sigma$ combined range of the Higgs boson
mass from Ref.\cite{giardino}.} \label{observability}
\end{center}
\end{figure}

\begin{table}[htb]
\caption{Branching ratio for top quark decay $t\rightarrow cH$ in
different models.}
\begin{center}
\begin{tabular}{|c|c|c|c|c|c|c|c|c|}
\hline
   & SM & QS & 2HDM & FC 2HDM & MSSM & $R \!\!\!\!\!\!  \not \quad$ SUSY
   &SUSY-QCD&LHT
   \\
   \hline
$t \to c H$ & $3 \times 10^{-15}$ & $4.1 \times 10^{-5}$
  & $1.5 \times 10^{-3}$ & $\sim 10^{-5}$
  & $10^{-5}$ & $\sim 10^{-6}$&$\sim10^{-5}$&$\sim10^{-5}$\\\hline
\end{tabular}
\end{center}
\end{table}
At the LHC, the dominant background of the search for $t\rightarrow
cH$ is the final state of $4j/3b\ell\nu$ coming from top quark pair
production: $pp \to t\bar{t} \to b \ell^+ \nu \bar{b}\bar{c}s+X$ or
$pp \to t\bar{t} \to b \ell^- \bar{\nu} bc\bar{s}+X$, where a
$c$-jet is mis-identified as a $b$-jet. The mis-tagged probability
of a $c$-jet as a $b$-jet is approximately 10\% reported by the
ATLAS and CMS. In order to investigate the observability of
$t\rightarrow cH$ for case III in the LHT model, we use the monte
carlo simulation results in Ref.\cite{hou} and plot 3$\sigma$ and
5$\sigma$ contours of the hadronic cross sections $pp \to t\bar{t}
\to b\ell \nu b\bar{b}j$ in Fig.\ref{observability} for
$\sqrt{s}=8,14$ TeV. We use the next-to-leading order value of
$t\bar{t}$ production rate in the calculation. Since the branching
ratio of $t \to cH$ is sensitive to the third generation mirror
quark mass, we take $M_{3}=1000 \rm ~GeV, 2000 \rm ~GeV, 3000 \rm
~GeV$ for example, where we set the scale $f=500\rm ~GeV$, the
mixing parameter $x_{L}=0.1$ and the first two generation mirror
quark masses $M_{12}=750 \rm~GeV$.

On the left panel of Fig.\ref{observability}, we can see that when
$M_3>2.2$ TeV, $t\to cH$ can reach $3\sigma$ sensitivity at 8 TeV
LHC with luminosity $\cal{L}$ $=20 fb^{-1}$. But on the middle and
right panel of Fig.\ref{observability}, we can find that the 14 TeV
LHC has the ability to probe the value of $M_3$ as low as 2.1(1.5)
TeV at $5\sigma$ level when $\cal{L}$ $=30(300) fb^{-1}$. Therefore,
we can infer that the precise measurement of $t\bar{t}$ production
can give a strong constraint on the parameters space of the LHT
model.

\begin{figure}[htbp]
\begin{center}
\scalebox{0.7}{\epsfig{file=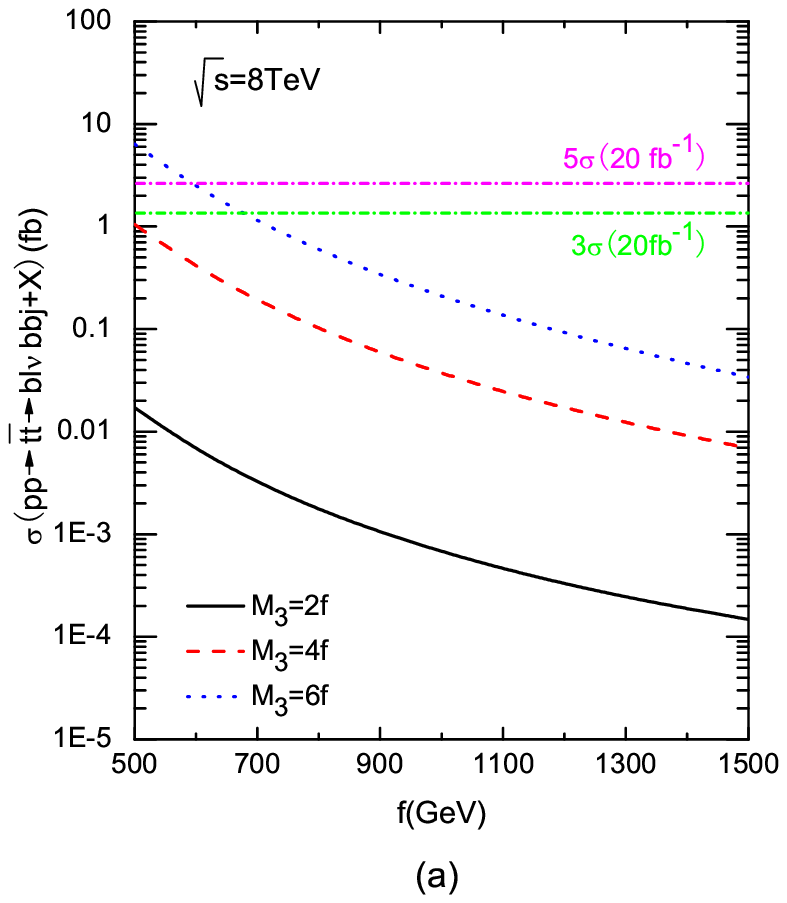}}\hspace{-0.5cm}
\scalebox{0.7}{\epsfig{file=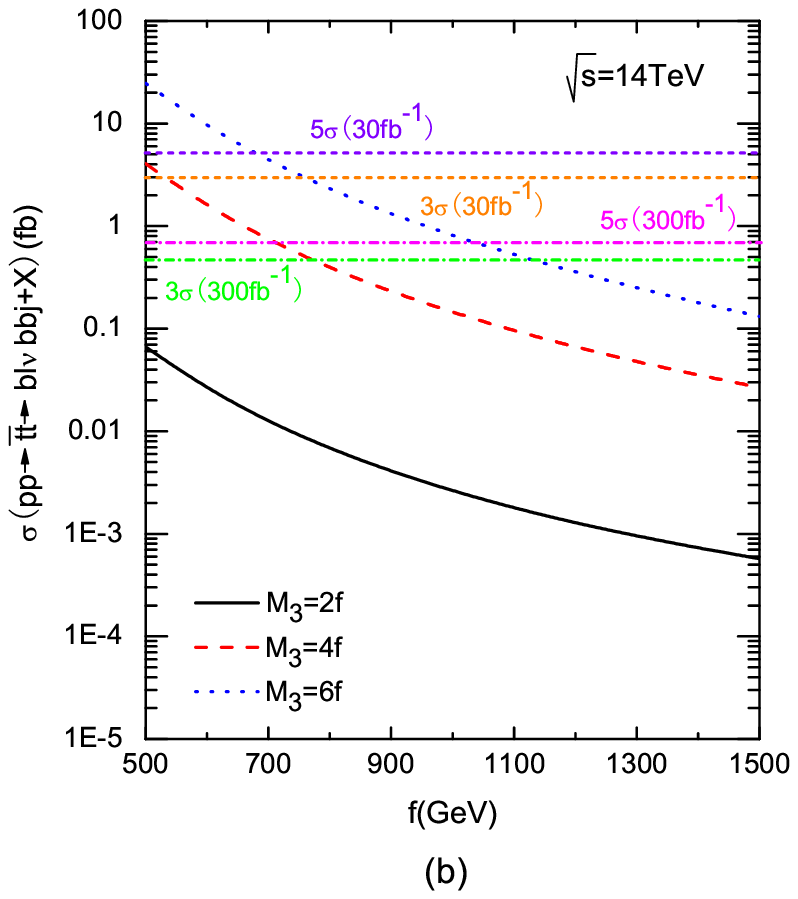}}\vspace{-1cm} \caption{The
observability of $t \to cH$ as a function of the scale $f$ for three
cases in the LHT model through the production of $pp\rightarrow
t\bar{t}\rightarrow tcH\rightarrow bl\nu cb\bar{b}+X$ at the LHC
with $\sqrt{s}=8$ TeV and $\sqrt{s}=14$ TeV.} \label{observabilityf}
\end{center}
\end{figure}

In Fig.4, we show the observability of $t \to cH$ as a function of
the scale $f$ for case III in the LHT model through the production
of $pp\rightarrow t\bar{t}\rightarrow tcH\rightarrow bl\nu
cb\bar{b}+X$ at the LHC with $\sqrt{s}=8$ TeV and $\sqrt{s}=14$ TeV.
We use the monte carlo simulation results and the next-to-leading
order value of $t\bar{t}$ production rate as above. The relevant
parameters are taken as follows: $x_{L}=0.1$, $M_{12}=1.5f$,
$m_{h}=125$ GeV. Based on the same consideration, we take the third
generation mirror quark mass $M_{3}=2f, 4f, 6f$ for example. We can
see that the favorable observability comes from the region with the
low $f$ and the large mass splitting $(M_{3}-M_{12})$, which is
consistent with the preceding analysis.

\section{Branching ratio for $t\rightarrow cg(\gamma)H$ in the LHT model}
\begin{figure}[htbp]
\scalebox{0.45}{\epsfig{file=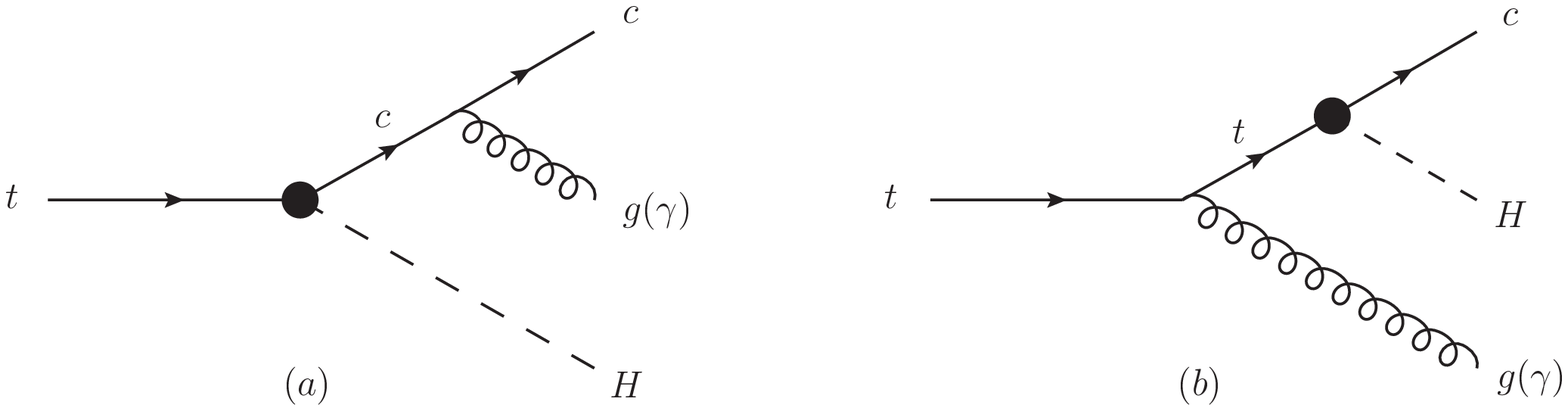}}
\scalebox{0.4}{\epsfig{file=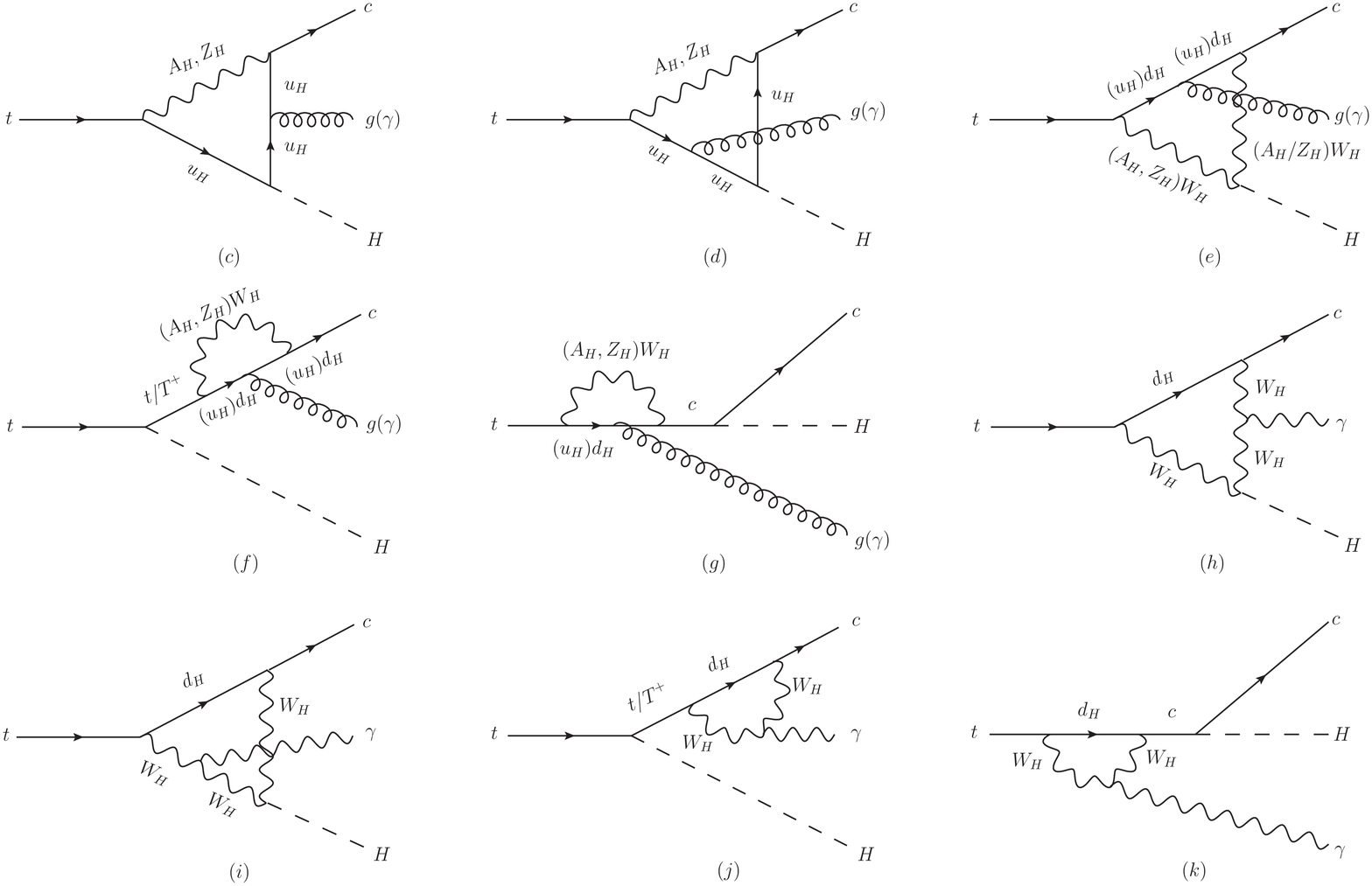}}\caption{Feynman diagrams of
the LHT one-loop correction to $t\rightarrow cg(\gamma)H$ in the
unitary gauge.}
\end{figure}
\noindent

In this section, we calculate the branching ratio of $t\rightarrow
cg(\gamma)H$ in the LHT model. These processes can also be
considered as part of the next-to-leading order QCD(QED) corrections
to $t\rightarrow cH$. The relevant Feynman diagrams of the process
$t\rightarrow cg(\gamma)H$ in unitary gauge are shown in Fig.5,
where the black dot represent the loop-induced $tcH$ vertex as shown
in Fig.1. In the numerical calculations, we take the same parameters
and cases as the decay process $t\rightarrow cH$ and impose the
kinematical cuts on the final massless states to avoid the
singularity.

\begin{figure}[htbp]
\begin{center}
\scalebox{0.7}{\epsfig{file=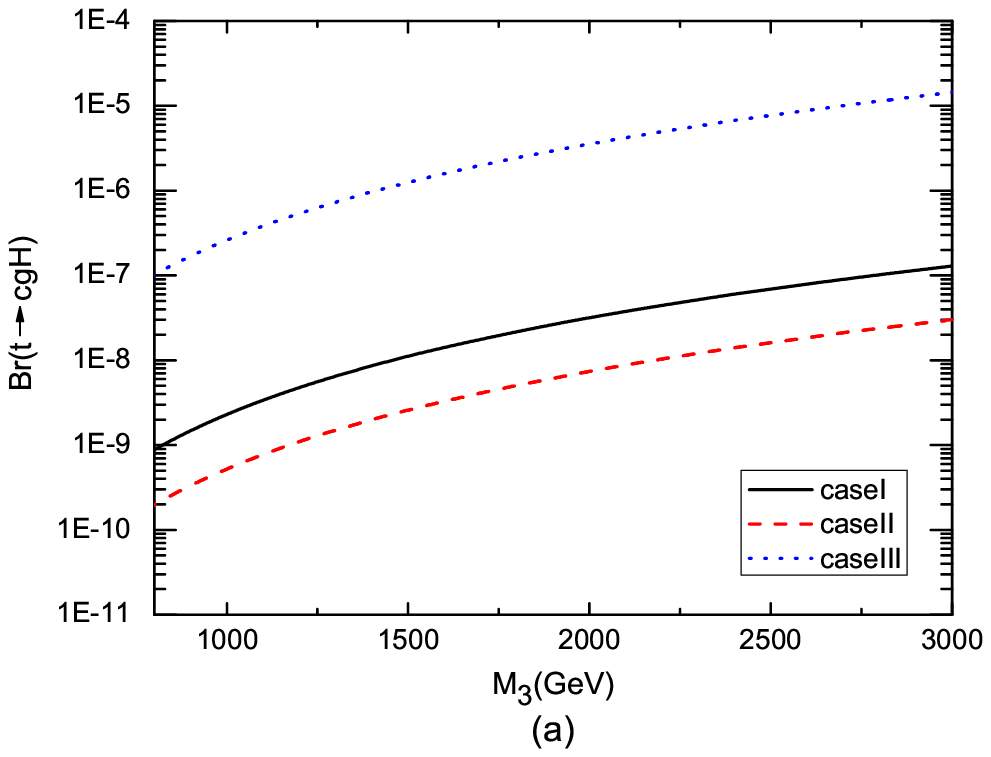}}\vspace{-1cm}\hspace{-1cm}
\scalebox{0.7}{\epsfig{file=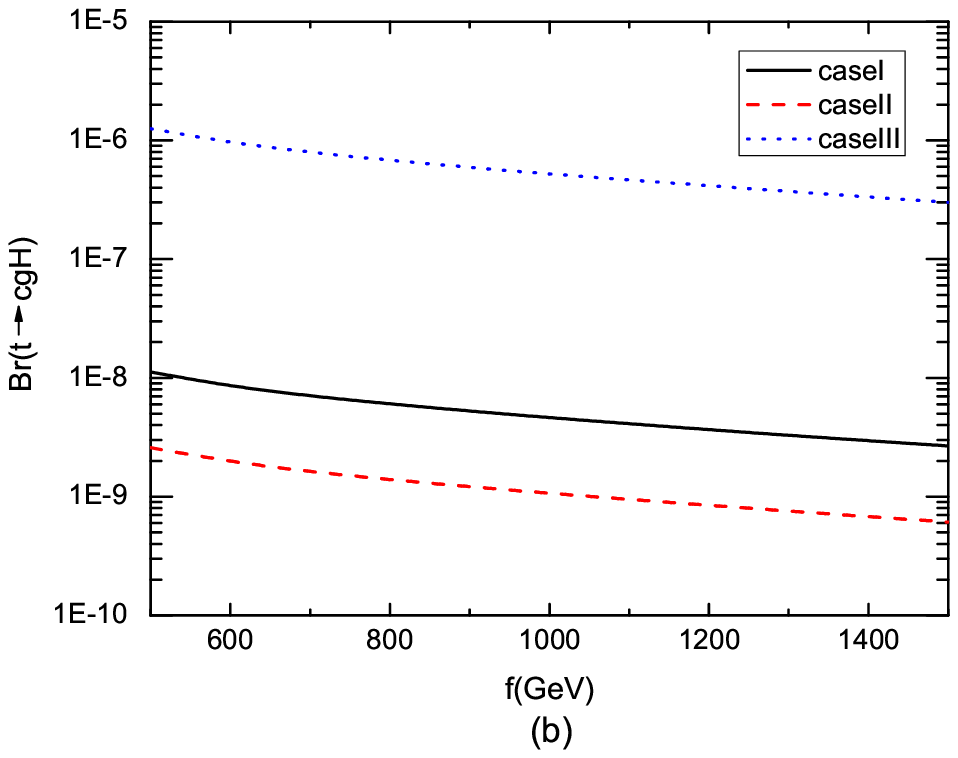}} \caption{Branching ratios
for $t\rightarrow cgH$ as a function of $M_{3}$ (a) and $f$ (b) in
three cases, respectively.}
\end{center}
\end{figure}

\begin{figure}[htbp]
\begin{center}
\scalebox{0.7}{\epsfig{file=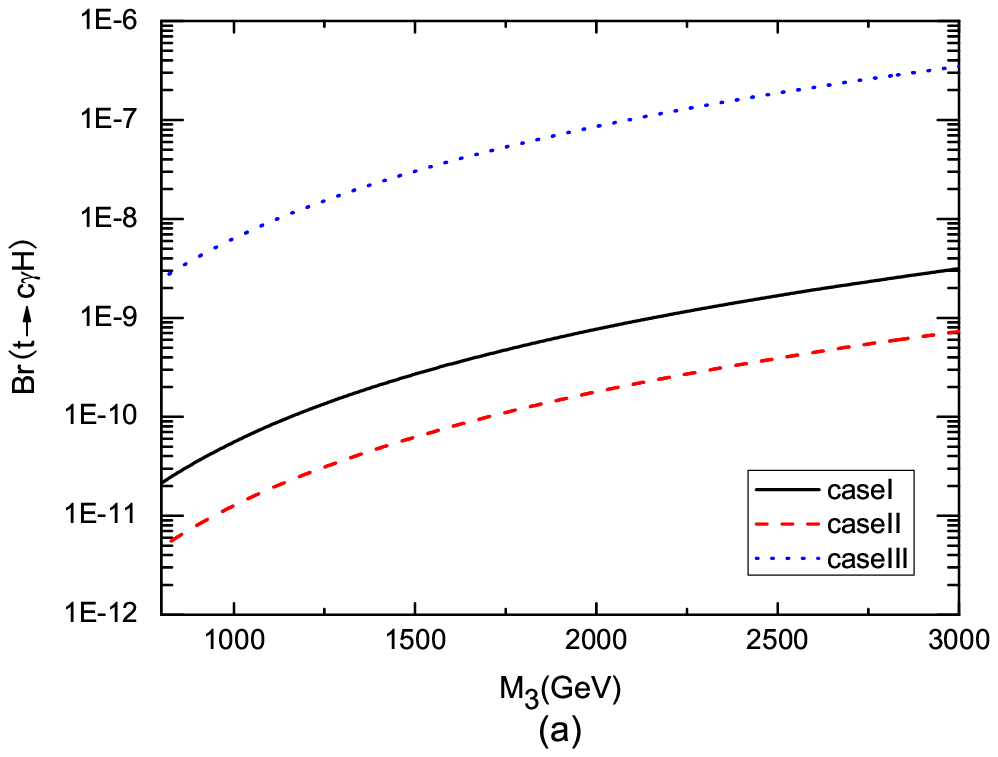}}\vspace{-1cm}\hspace{-1cm}
\scalebox{0.7}{\epsfig{file=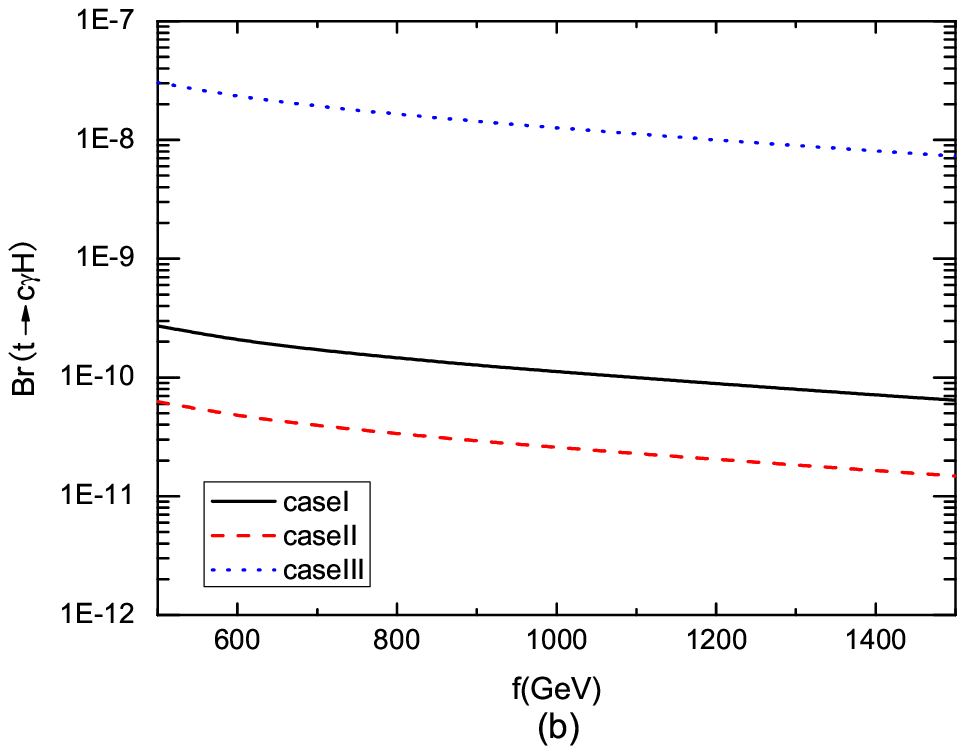}} \caption{Branching ratios
for $t\rightarrow c\gamma H$ as a function of $M_{3}$ (a) and $f$
(b) in three cases, respectively.}
\end{center}
\end{figure}

In Fig.6(a), we show the dependance of the branching ratio of
$t\rightarrow cgH$ decay process on the third generation mirror
quark mass $M_{3}$. We can see that the branching ratio of
$t\rightarrow cgH$ increases with $M_{3}$ increasing, the largest
branching ratio comes from the case III and the maximum value can
reach $1.4\times 10^{-5}$. In Fig.6(b), we show the dependance of
the branching ratio of $t\rightarrow cgH$ decay process on the scale
$f$. We can see that the correction of the LHT model decouples with
the scale $f$ increasing.

In Fig.7, we show the dependance of the branching ratio of
$t\rightarrow c\gamma H$ decay process on the third generation
mirror quark mass $M_{3}$(a) and the scale $f$(b). We can see that
the decay process $t\rightarrow c\gamma H$ has the similar behaviors
as the decay process $t\rightarrow cgH$. The maximum value of the
branching ratio can reach $3.5\times 10^{-7}$.

\section{summary}

\noindent

In this paper, we calculated the top quark FCNC decay $t\rightarrow
cH$ and $t\rightarrow cg(\gamma)H$ in the unitary gauge in the LHT
model. We found that the branching ratio for $t\rightarrow cH$ and
$t\rightarrow cg(\gamma)H$ can respectively reach $5.8\times
10^{-5}$ and $1.4\times 10^{-5}$($3.5\times 10^{-7}$) in the allowed
parameter space. When the mirror fermion mass $M_3 > 2.2$ TeV and
the cut-off scale $f=500$ GeV, $t\to cH$ can reach $3\sigma$
sensitivity at 8 TeV LHC with luminosity $\cal{L}$ $=20 fb^{-1}$. We
also noted that the 14 TeV LHC has the potential to observe this
channel at $5\sigma$ sensitivity level for $M_3 = 2.1(1.5)$ TeV when
$\cal{L}$ $=30(300) fb^{-1}$. Therefore, we can see that
$t\rightarrow cH$ may be used to test the LHT model at the LHC.

\section*{Acknowledgement}
We appreciate the helpful discussions with Jinmin Yang, Junjie Cao,
Lei Wang and Lei Wu. Ning Liu would like to thank Dr Archil
Kobakhidze for his warm hospitality in Sydney node of CoEPP in
Australia. This work is supported by the National Natural Science
Foundation of China (NNSFC) under grant Nos.11275057, 11305049,
11347140 and Specialized Research Fund for the Doctoral Program of
Higher Education under Grant No.20134104120002.

\end{document}